# Google Scholar Metrics evolution: an analysis according to languages

**Enrique Orduña-Malea**[1,*] **and Emilio Delgado López-Cózar**[2]

[1]EC3 Research Group, Universidad Politécnica de Valencia. Camino de Vera s/n, Valencia 46022, Spain.
[2]EC3 Research Group, Universidad de Granada, 18071 Granada, Spain
*e-mail: enorma@upv.es

**Abstract** In November 2012 the Google Scholar Metrics (GSM) journal rankings were updated, making it possible to compare bibliometric indicators in the 10 languages indexed – and their stability – with the April 2012 version. The h-index and h-5 median of 1,000 journals were analysed, comparing their averages, maximum and minimum values and the correlation coefficient within rankings. The bibliometric figures grew significantly. In just seven and a half months the h-index of the journals increased by 15% and the median h-index by 17%. This growth was observed for all the bibliometric indicators analysed and for practically every journal. However, we found significant differences in growth rates depending on the language in which the journal is published. Moreover, the journal rankings seem to be stable between April and November, reinforcing the credibility of the data held by Google Scholar and the reliability of the GSM journal rankings, despite the uncontrolled growth of Google Scholar. Based on the findings of this study we suggest, firstly, that Google should upgrade its rankings at least semi-annually and, secondly, that the results should be displayed in each ranking proportionally to the number of journals indexed by language.

## 1. Introduction

Google Scholar Metrics (GSM)[1] is a free product launched in April 2012 by Google that provides bibliometric information on a wide range of scholarly journals,[2] as well as other published material, such as conference articles and repositories.

The selection of journals is based on publications that are indexed in Google Scholar (excluding dissertations, books and patents), have published at least 100 articles over a period of five years, and have received at least one citation in that time.[3] These criteria represent an effort (albeit occasionally automatic and crude) to filter raw data from the journals indexed in Google Scholar (GS) and which, together with Google Scholar Citations (Google's product for the creation of researcher profiles) makes up Google's array of academic information tools today.

The features of this new product, its services, limitations and potential uses as a tool for the evaluation of academic activity, have been analysed in previous studies (Delgado-López-Cózar and Cabezas-Clavijo 2013), in which certain strengths (ease of use, coverage, free-of-charge, etc.) are highlighted, which are more an asset of the features of Google technology than of the product itself, which favour its use as a complementary source. Furthermore, the draw of the Google brand and the convenience of not having to install additional software point to a potential large-scale use of this product by various different stakeholders from the academic world (Jacsó 2012).

These previous analyses conclude that the GSM tool is still a product in its infancy with multiple limitations and errors, so that its use in evaluation processes is discouraged at present (Delgado-López-Cózar and Cabezas-Clavijo 2012). The recent appearance of this product means that there are many aspects to be investigated, in particular, those related to the reliability and validity of journal rankings based on the h-index. One of the aspects that is yet to be tested, given the rapid growth of GS, is the stability and variation of indicators over time.

In November 2012, GSM updated journal rankings published in the first version of April 2012 (fig. 1), extending the citation window up to 15 November of that year (fig. 2). Although there is no information on the product's website about any policy of regular content updates, this second version in the same year made it possible to take a quick and easy look at changes in the various journal rankings and, especially, at the growth and stability of the rankings. It therefore admitted, for the first time, longitudinal studies.[4]

**Figure. 1. Google Scholar Metrics (April 2012 version)**
**Figure. 2. Google Scholar Metrics (November 2012 version)**

Since GSM is fed by GS data, a study of the former must be contextualised against the of the latter. In fact, any study on the evolution of GS (or any other web data source) must also be aligned with studies related to the dynamism and evolution of search engines and, in general terms, of the Web (Brewington 2000), based on research into the durability and stability of online resources over time as well as the control variables.

In this regard, web resources can remain stable, change, disappear and reappear (intermittence effect), modified or not (which explains how the number of records in GS can fall over time). Studies worth mentioning in this area are those by Koehler (2002; 2004), Cho and García-Molina (2003) and Fetterly et al. (2003). In academic study environments, mention should also be made of the research by Ortega et al. (2006) and Payne and Thelwall (2007), among others.

If there are few longitudinal studies on the evolution of GS over time, those focused on measuring changes in various bibliometric values (h-index, total number of citations, etc.), as calculated from this database, are virtually nonexistent; as are comparisons with the rankings offered by traditional bibliometric journal evaluation systems (WoS and Scopus).

Although studies focusing on the coverage of search engines have demonstrated high variability and irregularity in their academic content updates (Orduña-Malea et al. 2010), GS is more stable and it is estimated that its size increases every two weeks (Aguillo 2012). Even so, the growth of GS is subject to greater instability than that of WoS or Scopus, since it is more vulnerable to certain changes in the Google search engine. Orduña-Malea et al. (2009) illustrated this in their study of the monthly evolution in the size of Spanish public university websites in GS, identifying stable and slightly positive trends during the months studied, with the exception of occasional measurements in which Google coverage changes significantly. In any case, knowledge and understanding of the growth of GS is essential to properly interpret bibliometric indicators produced on consultation of its records.

Another interesting line of research is the analysis of the growth of the number of citations collected by GS over time. Kousha and Thelwall (2007) collected citations received by a sample of 880 articles (39 open access journals listed in WoS) from GS in two different time periods (October 2005 and January 2006), observing significant citation growth patterns, although this was dependent on the area of knowledge.

Chen (2010) compared citations obtained from GS for a set of articles from different databases with citations previously obtained by Neuhaus et al. (2006) for the same sample, revealing a significant increase in the coverage of GS.

Finally, Winter et al. (2013) also analysed the citations retrieved on GS, in this case for a chosen set of 56 articles in various fields of knowledge, and compared them with citations retrieved on WoS. The novelty of this research lies in the calculation, not only of current growth, but also of retroactive growth, i.e. it studied the citations retrieved for an item at two different moments in time ("x" and "y"), but considered, for both samples, only the citations received up to date "x". Thus, it is possible to ascertain the growth of GS in literature looked at retrospectively. The results show that retroactive growth for GS is substantially higher than for WoS (GS median of 170% versus 2% for WoS); current growth is also higher, although to a lesser extent (GS median 54% versus 41% for WoS).

Recently, Harzing (2013) performed another GS-based longitudinal study, in order to test the stability of the coverage of this database over time and its direct effect on the calculation of bibliometric indicators. In this case, the analysis focused on researchers, but on a fairly small sample (20 Nobel science laureates), and limited to only four disciplines (physics, chemistry, medicine, and mathematics). For his study, he collected the total citations received by these authors at three moments in time (April 2011, September 2011 and January 2012), and calculated the h-index for each author at each moment. The main outcome of this study is that the citations increase by a ratio of approximately 3% per month, although growth is unequal, depending on the area of knowledge. However, the growth of the h-index is moderate (except in chemistry, where the coverage grows more sharply), and furthermore, the average h-index in Scholar for each researcher in the areas of physics and medicine is very similar to calculations for WoS.

Harzing's study is extremely important because it indicates, in summary, that GS is growing rapidly but that the h-index continues to be more stable, which is precisely one of the advantages of using this indicator (Costas and Bordons, 2007). Moreover, it obtains similar values to classical bibliographic databases (Scopus and WoS).

The GSM update is therefore an opportunity to expand on and confirm Harzing's findings in a larger sample, and for journals. In addition, GSM has a special feature whereby it provides a ranking of journals based on the 10 most representative languages in the world, a resource unusual in bibliometrics (Delgado López-Cózar and Cabezas-Clavijo, 2013). Therefore, the study of its evolution between the April and November versions may be contextualised in relation to language, a novel aspect in this type of study.

## 1.1. Objectives

Given the recent apparition of GSM, no other longitudinal study of this nature has been published to date. This paper therefore aims to answer the following questions:

a) Have there been changes in the values of the bibliometric indicators adopted by Google Scholar Metrics between April and November? Can a journal's h-index

Google Scholar Metrics evolution: an analysis according to languageschange in just seven months and a half? And if so, what is the volume and size of the variation?
b) Are there significant differences in the h-index for rankings by different languages?
c) Do these changes affect the positions of the journals in the rankings? In other words, is there stability in the rankings?

Thus, the specific objectives of this study are:

1. Measure the differences in the bibliometric indicators of the 10 journal rankings provided by Google Scholar Metrics, according to language, between April and November 2012.
2. Compare the stability of the rankings at these two moments in time.
3. Contextualise the results according to coverage and overall growth of GS as the GSM data source.

## 2. Methodology

In order to meet the objectives and to respond to the questions raised above, we designed a prospective, longitudinal, descriptive analysis of GSM.

For this purpose, we took a sample of the 1,000 journals listed in the rankings provided by GSM in April and November 2012 by language, which were as follows: English, Chinese, Portuguese, German, Spanish, French, Korean, Japanese, Dutch and Italian.

For each of the journals, and in each of the languages, the h-index and the median h-index (h-median) were obtained for both April and November. The data were then transferred to a spreadsheet where h-index and h-median averages were calculated, as well as maximum and minimum values, both overall and per language. Finally, the Spearman correlation coefficient of the journal rankings was calculated by language for the two versions (April and November). This process took place during the month of December 2012.

The bibliometric indicators were retrieved just as they were provided by GSM, in this case with a citation window of 5 years (h5-index and h5-median):[5]

- The h-index of a publication is the largest number h such that at least h articles in that publication were cited at least h times each.
- The h-median of a publication is the median of the citation counts in its h-core (the articles that the h-index is based on).
- Finally, the h5-index and h5-median of a publication are, respectively, the h-index and h-median of only those of its articles that were published in the last five complete calendar years.

It should be noted that GSM does not keep a file of journal positions and data from the previous version. In other words, the November 2012 update deleted the information from the first April 2012 version (Delgado López-Cózar and Cabezas-Clavijo, 2013). For the purposes of this study, data from the first version were captured prior to their deletion, giving added value to the analysis and comparison with November data.[6]

Since the growth of GSM relies on GS, we then conducted an additional experiment in which we collected data on GS coverage (measured in number of records), in order to contextualise the evolution of GSM. To match this with the GSM rankings by language, data on overall size were collected from the geographic domains of the main countries whose official language is one of the 10 considered by GSM.

This methodology for calculating the size of GS through cybermetric indicators has already been employed by Aguillo (2012). To compare the growth of GS with the other bibliometric databases, the number of records for the same countries was retrieved from WoS and Scopus. The data for the three databases were collected on a weekly basis during the month of March 2013 (four samples) and in November 2012 (with the release of the second version of GSM). Both total records and records from 2003 onwards were retrieved, to ascertain the rate of growth for contemporary literature.

The countries considered and queries performed for each of the three databases are shown in Table 1. In the case of the United States, GS measurement was performed by adding the results to the domains .edu, .gov, .mil, and .us. Despite the limitations of this procedure (overlapping records and use of the domain .edu in other countries), this is the only process for retrieving geographic web data for the US.

**Table 1. Countries, sources and queries performed**

| COUNTRY | SCHOLAR | SCOPUS | WoS |
|---|---|---|---|
| **USA** | site:edu +site:gov +site:mil +site:us | AFFILCOUNTRY(united states) | AD= (USA) OR AD= (united states) |
| **UK** | site:uk | AFFILCOUNTRY(united kingdom) | AD= (uk) OR AD= (england) OR AD= (united kingdom) |
| **Italy** | site:it | AFFILCOUNTRY(italy) | AD= (italy) |
| **France** | site:fr | AFFILCOUNTRY(france) | AD= (france) |
| **Germany** | site:de | AFFILCOUNTRY(germany) | AD= (germany) |
| **Netherlands** | site:nl | AFFILCOUNTRY(netherlands) | AD= (netherlands) |
| **Spain** | site:es | AFFILCOUNTRY(spain) | AD= (spain) |
| **Brazil** | site:br | AFFILCOUNTRY(brazil) | AD= (brazil) |
| **Japan** | site:jp | AFFILCOUNTRY(japan) | AD= (japan) |
| **Korea** | site:kr | AFFILCOUNTRY(south korea) | AD= (south korea) |
| **China** | site:cn | AFFILCOUNTRY(china) | AD= (china) |

This procedure is not intended to consider the GS / WoS-Scopus isomorphism, since cybermetric indicators applied to GS retrieve the records deposited in the domain of each country, which will not necessarily be the output of that country (although the majority undoubtedly is). In addition, there are countries that have not been considered for the most important languages, like English (Australia), Spanish (all of South America), German (Austria) or French (Canada, Africa, etc.). However, this simple method is effective for determining, in a simple exploratory manner, the growth of GS as backdrop for GSM, and how it compares with WoS and Scopus.

## 3. Results

First, results are provided for the growth of the two main bibliometric indicators (h-index and h-median) between April and November 2012. Then detailed information is Google Scholar Metrics evolution: an analysis according to languagesgiven on growth data according to the GSM language ranking and ranking correlations retrieved from both version for each of the languages analysed.

Finally, GS, Scopus and WoS growth data are given (total and 2003 onwards) for the selected countries in order to compare and contextualise journal rankings, according to language, that have been the object of previous studies.

### 3.1. Growth of indicators: h-index and h-median

From the analysis of the 1,000 scholarly journals displayed in the 10 GSM rankings, significant growth is observed in the bibliometric values adopted by Google to measure academic impact. Table 2 shows the results for the h-index and h-median (average, maximum and minimum values in both cases) for April and November .

**Table 2. Bibliometric indicators of journal rankings published by GSM**

| Indicators | April 2012 | November 2012 | Growth rate (%) |
|---|---|---|---|
| h-index (average) | 20 | 23 | 15.4 |
| h-index (maximum) | 47 | 58 | 16.5 |
| h-index (mimimum) | 15 | 18 | 19.0 |
| h-median (average) | 29 | 34 | 17.0 |
| h-median (maximum) | 72 | 85 | 18.9 |
| h-median (minimum) | 17 | 23 | 24.9 |

As seen in Table 2, the h-index average for the 1,000 journals has grown from 20 to 23 (15.4%) in less than eight months, while the h-median has a slightly higher growth rate (17%). This increase in the indicators also occurs in the other parameters considered (minimum and maximum values of both the h-index and the h-median). There is, however, a notable increase in the minimum values, 24.9% in the case of the h-median, a figure considered very significant given the limited time period in which it is produced.

### 3.2. Ranking according to language: differences and growth

After analysing overall indicator values, Table 3 presents the results obtained for the journal sets published in each language. Here it is apparent how the bibliometric indicators for journals in English have significantly higher values than those for the other languages, five times the average h-index values obtained for the language ranked second (Chinese), while Portuguese and Spanish occupy third and fourth place in both samples, a certain distance behind Chinese.

**Table 3. Bibliometric indicators of journal rankings published by GSM according to language**

| LANGUAGE | H-INDEX | | | H-MEDIAN | | |
|---|---|---|---|---|---|---|
| | April | November | Growth rate (%) | April | November | Growth rate (%) |
| English | 121 | 138 | 14.0 | 174 | 199 | 14.4 |
| Chinese | 23 | 27 | 17.4 | 31 | 38 | 22.6 |
| Portuguese | 14 | 17 | 21.4 | 19 | 24 | 26.3 |
| Spanish | 11 | 13 | 18.2 | 16 | 18 | 12.5 |
| German | 10 | 12 | 20.0 | 15 | 18 | 20.0 |
| French | 7 | 8 | 14.3 | 10 | 13 | 30.0 |
| Korean | 5 | 6 | 20.0 | 7 | 9 | 28.6 |
| Japanese | 5 | 5 | 0.0 | 6 | 7 | 16.7 |
| Italian | 3 | 4 | 33.3 | 5 | 7 | 40.0 |
| Dutch | 2 | 2 | 0.0 | 5 | 4 | -20.0 |
| AVERAGE | 20 | 23 | 15.4 | 29 | 34 | 17.0 |

The fact that Chinese journals prominently occupy second place in the h-index average journal ranking is not surprising, given the large amount of current scholarly production in this language (as discussed later). Conversely, the high position of journals in Portuguese is striking, higher than for those published in Spanish (the differences remain constant in November), considering that the size of the Spanish-speaking scientific community is greater than the Portuguese. Also of note are the low values for French, which ranks sixth.

As for the differences between April and November, there is an increase in h-index across nearly all the rankings by language. Growth rates are highest for journals in Italian (33.3%) and Portuguese (21.4%), while the lowest are for Japanese and Dutch, which are the only languages with a constant h-index (5 and 2 respectively).

In any case, these growth rates should be approached with some caution, due to the nonlinearity of the growth in the h-index. The h-index for Italian-language journals increases from 3 to 4 (representing a 33.3% growth). The value of the h-index for other languages has also increased by one point, for example Korean (from 5 to 6) and French (from 7 to 8), yet their growth rates are lower since the index is higher (20% and 14% respectively). This is especially significant as it is more difficult for an h-index to increase from 7 to 8 than it is to go from 3 to 4.

If we focus on this issue, Table 3 then shows how the journals in languages that already had high values in April are those that most increase their h-index values in November (in whole integers): English increases 7 points, Chinese 4, Portuguese 3, and Spanish and German 2.

### 3.3. Correlation analysis: stability of the rankings

From the point of view of the bibliometric evaluation of journals, one of the key aspects in determining the reliability and validity of the GSM rankings is their stability over time. In this respect, both the coverage and growth of GS, on the one hand, and the robustness and progressiveness of the h-index, on the other, are established as key variables of analysis, as noted in the introduction.

Table 4 shows the correlation coefficients (Spearman) between GSM journal rankings in April and November for the following languages:

Table 4. Correlation coefficients (Spearman) between GSM editions (April and November)

| LANGUAGE | H-INDEX | H-MEDIAN |
|---|---|---|
| **English** | 0.99 | 0.99 |
| **Portuguese** | 0.97 | 0.95 |
| **Spanish** | 0.96 | 0.90 |
| **Chinese** | 0.95 | 0.94 |
| **German** | 0.95 | 0.94 |
| **French** | 0.92 | 0.85 |
| **Korean** | 0.88 | 0.63 |
| **Japanese** | 0.89 | 0.76 |
| **Italian** | 0.85 | 0.58 |
| **Dutch** | 0.73 | 0.87 |

As the data in Table 4 show, there is a very high correlation between the journal rankings of April and November. They also show how this correlation coefficient is slightly dependent on the language of the journals and their h-index, i.e. the correlation decreases practically to the same extent that the values of the h-index are of a lesser scale. This is a logical trend because, as discussed above, the h-index is more sensitive to errors the lower its value, becoming more robust the higher the value. For this reason the values achieved for English (.99), Portuguese (.97) and Spanish (.96) are so high. This fact clearly indicates that there is little variation in the positions of the journals.

### 3.4. Database growth: Scholar, WoS and Scopus

Finally, this section looks at the growth data for GS, as the GSM data source, and compares these data with those of Scopus and WoS.

Table 5 shows the statistical range obtained from weekly measurement of records in the three databases. The complete raw data for each country and database are available in the supplementary material for consultation.

The data correspond to both the total number of records and those obtained only from 2003 onwards. Moreover, each of the languages included in GSM is represented by a country, except in the case of English, represented both the UK and the United States.

In the case of GS, there are positive ranges (i.e. growth during the four weeks of measurement) for China (8,100,000 more records) and Italy (55,000). In the case of the United States, despite the fact that it also obtained a very high positive range (106,300 more records), the data must be treated with some caution, as they are the sum of four domains (.edu, .mil, .us, .gov), which means that there may be a high degree of overlapping. In addition, the .edu domain is also used in other countries, so the GS figures relating to web indicators for the United States should be taken only as a rough indication.

**Table 5. Statistical range (total and recent) of records retrieved in GS, Scopus and WoS**

| COUNTRIES | SCHOLAR | | SCOPUS | | WoS | |
|---|---|---|---|---|---|---|
| | R (tot) | R (recent) | R (tot) | R (recent) | R (tot) | R (recent) |
| **UK** | -2,630,000 | -518,000 | 10,107 | 9,802 | 8,671 | 8,656 |
| **Italy** | 4,000 | 55,000 | 5,752 | 5,637 | 5,452 | 5,399 |
| **France** | -240,000 | -160,000 | 7,154 | 6,988 | 5,762 | 5,753 |
| **Germany** | -10,000 | -41,000 | 9,528 | 9,324 | 8,821 | 377,497 |
| **Netherlands** | 27,000 | -6,000 | 3,248 | 3,200 | 2,991 | 3,003 |
| **Spain** | -250,000 | 10,000 | 4,886 | 4,807 | 4,875 | 4,868 |
| **Brazil** | -10,000 | 0 | 3,807 | 3,777 | 3,052 | 3,034 |
| **Japan** | 0 | 10,000 | 7,120 | 7,020 | 6,651 | 6,640 |
| **Korea** | 20,000 | -92,000 | 4,399 | 4,364 | 6,159 | 3,131 |
| **China** | 9,100,000 | 8,100,000 | 24,520 | 24,495 | 14,499 | 14,486 |
| **USA** | 1,227,000 | 106,300 | 35,811 | 33,280 | 35,249 | 32,861 |

For recent literature (from 2003 onwards), Netherlands (6,000 fewer records), Germany (41,000), South Korea (92,000), France (160,000) and, in particular, the UK (518,000) have a negative range. Finally, Brazil has inconsistent results (while recent literature remains the same, the total figure decreases significantly). In fact, recent data show certain inconsistencies in relation to total data for many countries, such as Italy (lower total growth than recent), Germany (greater recent decrease than total), Spain (total decrease and recent growth) and South Korea (total growth and recent decrease).

These results are consistent with the instability and limitations of GS in retrieving web data due to changes in Google coverage. For example, in the third sample (see supplementary materials) a fall in UK records may be observed (from 3,830,000 to 1,190,000, which helps to explain the final figures obtained).

Table 6 expands the GS results, comparing the data from the last sample (March 2013) with data collected just at the time of the appearance of the second version of the rankings (November 2012). It also adds the data calculated by Aguillo (2013), collected in August 2010.

Table 6. Google Scholar evolution according to geographic web domain

| COUNTRY | August 2010 | November 2012 | March 2013 |
|---|---|---|---|
| China | 7,520,000 | 18,900,000 | **30,700,000** |
| USA | 7,873,000 | 14,531,000 | **16,019,000** |
| Japan | 1,720,000 | 10,100,000 | **10,400,000** |
| France | 2820,000 | 4,430,000 | **4,210,000** |
| Spain | 907,000 | 3,140,000 | **2,990,000** |
| Brazil | 1,440,000 | 2,170,000 | **2,320,000** |
| Korea | 481,000 | 1,630,000 | **1,720,000** |
| UK | 430,000 | 4,230,000 | **1,200,000** |
| Germany | 684,000 | 1,060,000 | **1,030,000** |
| Italy | 308,000 | 723,000 | **798,000** |
| Netherlands | 219,000 | 717,000 | **774,000** |

The results show, on one hand, the high growth rate since the first data were reported (when the United States still surpassed China) up to 2013, also highlighting the high rate of growth in Japan. In fact, just as the figures for China and the United States bear relation to the data in Table 3, the low h-index values obtained for Japanese journals contrast with the Japanese web domain size, an aspect that should be studied in greater depth in the future.

Taking the last week of data collection as a reference, the countries with the highest total number of records are China (30,700,000 records), followed by the United States (16,019,000), Japan (10,400,000), France (4,210,000) and Spain (2,990,000). The weekly evolution of the number of records for these countries is shown in Figure 3, where we can see how, despite the observed data, growth is relatively stable, except for the weeks when there are updates; this is consistent with the results obtained previously by Aguillo (2012).

**Figure 3. Evolution of page count according to geographic domains in Google Scholar.**

The positions of the countries correspond approximately to the rankings provided in Table 3, with the exception of Portuguese and German. In any case, caution should be exercised with these results, as they evidently do not correspond to the actual growth data by country since every citizen or institution is free to choose the geographical or generic name when registering a domain name, although this process is indeed useful for determining geographic coverage in general.

Google Scholar Metrics evolution: an analysis according to languagesAs for the records retrieved from the other databases, the results are as expected: the number of records is lower, their growth rates are more stable and there are hardly any inconsistencies between total and recent results, with the exception of Germany, for which high figures were obtained for the number of recent records in WoS between the first (802,889) and the second sample (1,175,266).

Except for these last data – which are assumed to be the result of an error – after the United States, China is the country (among those analysed) with the greatest number of both recent and total records in Scopus and WoS. China's elevated results correspond to those displayed in Table 3, ranking according to language, and are especially significant given the high percentage of recent records in relation to the total (86.02% in Scopus; 82.33% in WoS), reflecting high productivity at present. In fact, South Korea, Japan and Brazil are the countries with the highest percentages of recent records.[7]

Figure 4, meanwhile, shows the weekly evolution of total records for China in the three databases analysed, in which differences in coverage and the staggered updates of GS may be observed. Moreover, the higher results in Scopus as compared to WOS are associated with the greater orientation of the former towards the Chinese publishing system (Leydesdorff 2012), a phenomenon which can also be seen clearly in Table 5.

**Figure. 4. Number of records for China (Scholar, Scopus & WoS)**

Finally, Table 7 shows a summary of the number of records (averaged weekly) and the monthly growth rate (from the first to the last weekly sample) by language and database, as well as total values (obtained from the sum of the number of records in all languages analysed).

Table 7. Weekly average size and monthly growth rate per source (Scholar, Scopus, WoS)

| COUNTRIES | GOOGLE SCHOLAR | | SCOPUS | | WOS | |
|---|---|---|---|---|---|---|
| | Weekly Size average | Monthly Growth rate | Weekly Size average | Monthly Growth rate | Weekly Size average | Monthly Growth rate |
| UK | 2,512,500 | -68.67 | 2,922,774 | 0.34 | 3,234,298 | 0.27 |
| Italy | 799,750 | 0.50 | 1,363,137 | 0.42 | 1,366,548 | 0.40 |
| France | 4,330,000 | -5.39 | 1,850,649 | 0.39 | 2,045,158 | 0.28 |
| Germany | 1,032,500 | -0.96 | 2,688,492 | 0.35 | 2,248,260 | 0.39 |
| Netherlands | 760,500 | 3.61 | 773,303 | 0.42 | 852,991 | 0.35 |
| Spain | 3,115,000 | -7.72 | 910,847 | 0.54 | 912,379 | 0.54 |
| Brazil | 2,325,000 | -0.43 | 519,823 | 0.73 | 498,969 | 0.61 |
| Japan | 10,400,000 | 0.00 | 2,571,141 | 0.27 | 2,641,581 | 0.25 |
| Korea | 1,712,500 | 1.18 | 610,588 | 0.72 | 593,955 | 1.04 |
| China | 26,150,000 | 42.13 | 2,823,879 | 0.87 | 2,001,858 | 0.73 |
| USA | 15,408,000 | 8.30 | 11,078,847 | 0.32 | 11,508,900 | 0.31 |
| TOTAL | 68,545,750 | 11.15 | 28,113,479 | 0.41 | 27,904,896 | 0.37 |

Although the variability of GS data is greater and changes are staggered, we can see how the average weekly number of records in GS is higher than in the other databases, except for UK, Italy, Germany and Netherlands, where the weekly average values are higher in both Scopus and WoS (due to the negative trends identified in GS for these countries). As for the total values, the prevalence of GS (due to the United States and, especially, to Japan and China) may clearly be observed, but the data must be interpreted with caution because of the methodological limitations already discussed regarding records extracted by geographic web domains.

With regard to growth rates, Table 7 shows, in summary, mismatches in GS data for some countries, both positive (China: 42.13%) and negative (UK: -68.67%). In any case, the values obtained for the entire set of languages clearly show a higher overall growth rate for GS in comparison with the other databases (GS: 11.5%; Scopus: 0.41%; WoS: 0.37%). Growth rates for individual countries must also be contextualised in relation to size in terms of the number of records (considerably higher in GS), to account for the effects of scale.

**4. Discussion and conclusions**

The data collected for this first longitudinal study of a representative sample of journals (1,000) in GSM, of very different origin (because they are published in 10 languages), show a significant increase in the h-index (15%) in a very short period of time (seven and a half months). These figures are higher than those reported by Harzing (2013) in his sample of the h-index of 20 Nobel laureates in physics, chemistry, medicine and economics (comprising 400 publications), in which the growth rate of the h-index was 7.3% (and 13.1% of the total citations). These lower values are justified by the fact that the period analysed in Harzing's study (April to September) is shorter than that applied in this study (April to November).

This remarkable increase in the GSM bibliometric indicators is evidently in keeping with, on the one hand, the spectacular growth of open access scholarly literature on the Internet and, secondly, the capacity of the GS search engine to index this academic literature immediately, which partly explains the staggered growth detected in the previous point.

The robot used by GS to automatically crawl the web immediately incorporates and processes all this information. In previous studies (Delgado López-Cózar and Cabezas-Clavijo, 2013), it was found that the indexing of a document and its citations does not take more than three days in the case of documents deposited in institutional or thematic academic repositories, and between a week and a month for the personal and institutional websites of scientists, research groups, departments, institutes and academic centres. Harzing (2013) showed a monthly citation growth rate of 3% in his study, double the increase in WoS. It is clear that GS is growing more and much faster than traditional databases. This growth in bibliometric values occurs for virtually every journal, regardless of the language in which it is published. However, significant differences in growth rates were detected, depending on the language in which the journal is published.

Another aspect revealed by this study is the marked differences in the size of the h-index for journals according to the publication language. The extremely high h-index of the English-language journals (ten times the values achieved by the journals in other languages, with the exception of Chinese where it is five times larger) is undoubtedly a faithful reflection of both the scale and size of the English-speaking academic community, the predominance of English as the lingua franca of academic communication, and the volume of scholarly production circulating on the Web.

An analysis of the growth of GS based on the web size of the main geographical top level domains (gTLD) shows, in particular, the dramatic increase in the number of items retrieved in the space of two years. Comparing the data collected by Aguillo in August 2010 with data collected in March 2013, the differences are illustrative enough to understand this phenomenon of English-language predominance. However, a particularly striking aspect is still the disproportionate position of Portuguese in relation to Spanish in terms of volume of data in the national domains and the h-index levels of the respective journals in GSM, as well as the results obtained for Japanese magazines.

The other important finding of this study is that, despite all the technical and methodological problems posed by GS as a source of information for academic evaluation (misidentification of documents and citations, lack of transparency in the selection of sources, deficiencies in the control and standardisation of records, etc.), the precise nature and true size of which we cannot determine (something that is, moreover, almost impossible given the universal nature of Google), the fact that its evolution is unpredictable, vulnerable to coverage of the commercial search engine, and different from the other bibliometric databases (i.e. WoS and Scopus), the stability of journal rankings between April and November 2012 is very high. This reinforces the credibility of the data handled by GS in the creation of GSM journal rankings, despite the fact that their size, coverage, updating and nature are, in principle, different from those provided by WoS and Scopus. Ultimately, journal rankings produced by GSM are sound and reliable.

Regardless of the data that this study has contributed to determining the size, nature and stability of bibliometric indicators produced by GS, conclusions may be drawn from this research which should lead to changes in the GSM rankings. These are:

1. The growth of bibliometric data demonstrates the need to update the rankings at least twice a year. This would be contrary to the usual practice of the bibliometric journal assessment industry, which updates its products annually. We would even go so far as to advise Google to turn its product into a fully dynamic information system, so that the rankings are updated instantly in the same way as the bibliometric author profiles provided by Google Scholar Citations.
2. Google Scholar Metrics is shown to have adopted a wrong policy in displaying only 100 journals in each ranking and, even worse, using the same threshold for rankings in the different languages. Such marked differences in the volume of indicators would require the results displayed in each ranking to be proportional to the number of journals indexed in every language. Our advice would be to use thresholds of 1%, 5% or 10% of the journals with the highest h-index by language.

## 5. Notes

[1] Google Scholar Metrics
http://scholar.google.com/citations?view_op=top_venues&hl=en (accessed 1 September 2013).
[2] http://googlescholar.blogspot.com.es/2012/04/googlescholar-metrics-for-publications.html (accessed 1 September 2013).
[3] http://scholar.google.com/intl/en/scholar/metrics.html (accessed 1 September 2013).
[4] In July 2013 the third version of GSM appeared, in which the citation window varied (2008-2012). For this reason this version is not taken into account in this longitudinal study. More information: http://arxiv.org/ftp/arxiv/papers/1307/1307.6941.pdf (accessed 1 September 2013).
[5] http://scholar.google.com/intl/en/scholar/metrics.html#metrics (accessed 1 September 2013).
[6] The raw data for the two GSM editions are available in Annex I of the supplementary material.
[7] The raw data for the weekly analysis of coverage in GS, Scopus and WoS are available in Annex II of the supplementary material.

# 6. References

Aguillo, Isidro F. (2012). Is Google Scholar useful for bibliometrics? A webometric analysis. *Scientometrics*, *91*(2), 343-351.

Brewington, B. E. & Cybenko, G. (2000). How dynamic is the Web?. *Computer Networks*, *33*(1-6), 257-276.

Cabezas-Clavijo, A. & Delgado López-Cózar, E. (2013). Google Scholar e índice h en biomedicina: la popularización de la evaluación bibliométrica. *Medicina intensiva*, *37*(5), 343-354.

Cho, Y. & Garcia-Molina, H. (2000). The evolution of the web and implications for an incremental crawler. *Proceedings of the 26th International Conference on Very Large Data Bases*, 200-209.

Costas, R. & Bordons, M. (2007). The h-index: Advantages, limitations and its relation with other bibliometric indicators at the micro level. *Journal of Informetrics, 1*(3), 193-203.

Delgado López-Cózar, E. & Cabezas-Clavijo, A. (2012). Google Scholar Metrics: an unreliable tool for assessing scientific journals. *El profesional de la información*, *21*(4), 419-427.

Delgado López-Cózar, E. & Cabezas-Clavijo, A. (2013). Ranking journals: could Google Scholar Metrics be an alternative to Journal Citation Reports and Scimago Journal Ranks. *Learned publishing*, *26*(2), 101-114.

Fetterly, D., Manasse, M., Najork, M. & Wiener, J. (2003). A large scale study of the evolution of web pages. *Proceedings of the Twelfth International Conference on World Wide Web*, 669-678.

Harzing, A-W. (2013). A preliminary test of Google Scholar as a source for citation data: a longitudinal study of Nobel prize winners. *Scientometrics*, *94*(3), 1057-1075.

Jacsó, P. (2012). Google Scholar Metrics for Publications – The software and content feature of a new open access bibliometric service. *Online Information Review*, *36*(4), 604-619.

Koehler, W. (2002). Web page change and persistence -4- year longitudinal web study. *Journal of the American Society for Information Science and Technology, 53*(2), 162-171.

Koehler, W. (2004). A longitudinal study of Web pages continued a consideration of document persistence. *Information Research*, *9*(2). Available at http://informationr.net/ir/9-2/paper174.html. (accessed 1 September 2013).

Kousha, K. & Thelwall, M. (2007). Google Scholar Citations and Google Web/URL citations: a multidiscipline exploratory analysis. *Journal of the American Society for Information Science and Technology*, *58*(7), 1055-1065.

Leydesdorff, L. (2012). World shares of publications of the USA, EU-27, and China compared and predicted using the new *Web of Science* interface versus *Scopus*. *El profesional de la información*, *21*(1), 43-49.

Orduña-Malea, E., Serrano-Cobos, J. & Lloret-Romero, N. (2009). Las universidades públicas españolas en Google Scholar: presencia y evolución de su publicación académica web. *El profesional de la información*, *18*(5), 493-500.

Orduña-Malea, E., Serrano-Cobos, J., Ontalba-Ruipérez, J-A. & Lloret-Romero, N. (2010). *Revista española de documentación científica*, *33*(2), 246-278.

Ortega, J. L., Aguillo, I. F. & Prieto, J. A. (2006). Longitudinal study of contents and elements in the scientific Web environment. *Journal of Information Science*, *32*(4), 344-351.

Payne, N. & Thelwall, M. (2007). A longitudinal study of academic webs: growth and stabilization. *Scientometrics*, *71*(3), 523-539.

Winter, Joost C.F. de, Zadpoor, Amir A. & Dodou, D. (2013). The expansion of Google Scholar versus Web of Science: a longitudinal study. *Scientometrics* [Online first]. DOI 10.1007/s11192-013-1089-2.